# How AI Is Reconfiguring Work in India

## A task-based framework for understanding augmentation, displacement, new roles and the changing structure of jobs

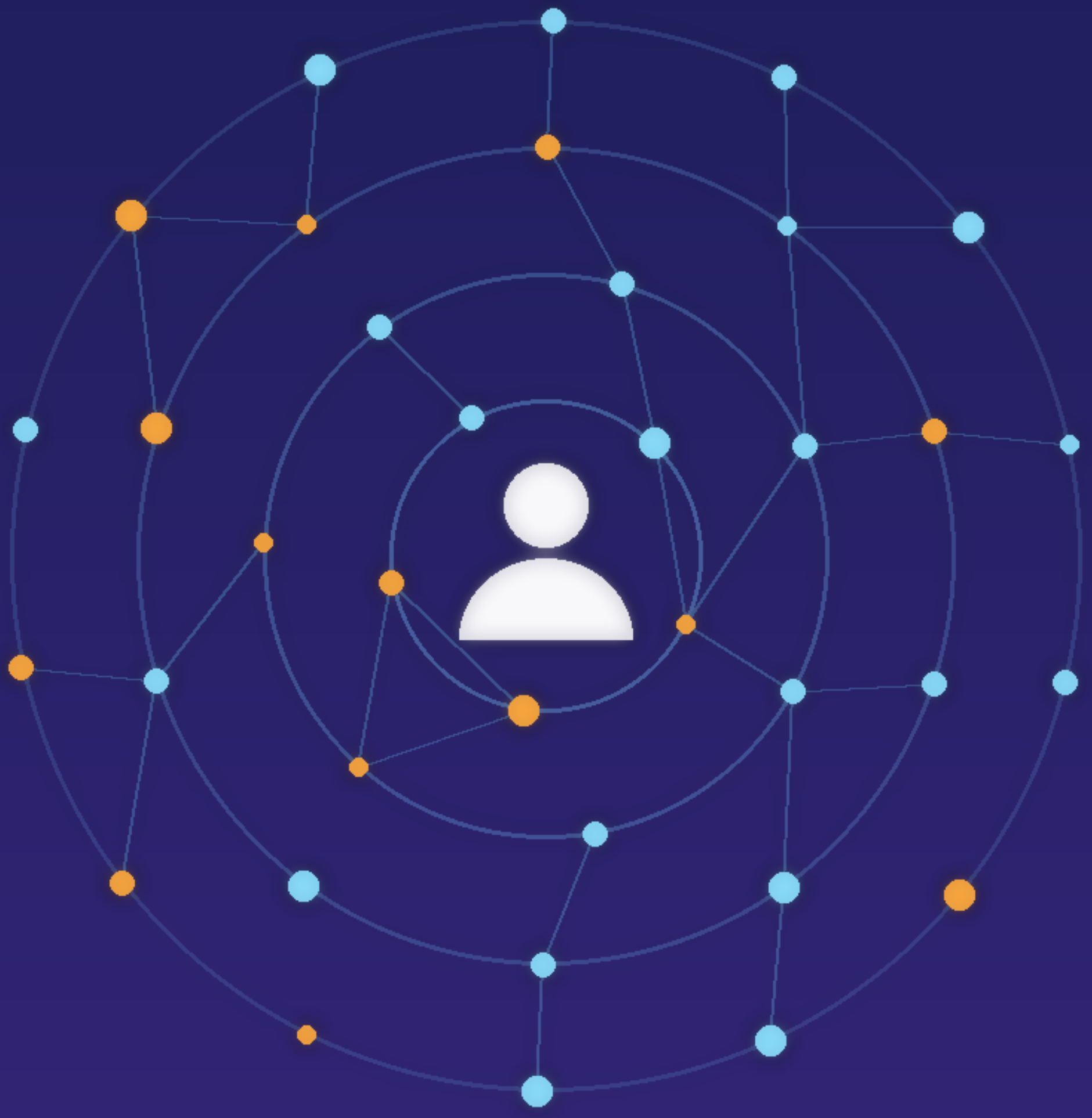

Centre for Responsible AI (CeRAI)
Wadhwani School of Data Science and AI (WSAI)
Indian Institute of Technology Madras

## Contributors

**Prof Balaraman Ravindran**
Head of CeRAI and WSAI, IIT Madras

**Ms VS Srividhya**
Executive Head, Srilatha and Prabhakar Raghavan Centre for Information & Society (CISoC), IIT Madras

**Mr Venkatraghavan KS**
Senior Advisor, CeRAI IIT Madras

**Dr Gokul S Krishnan**
Research Director, CeRAI IIT Madras

*The contributors acknowledge the use of Claude (Anthropic) for assistance in the preparation of this paper. All the content was reviewed, verified, and edited by the authors, who take full responsibility for the accuracy and integrity of the published work.*

## ABSTRACT

Public debate on AI and Indian employment tends to compress a complicated phenomenon into a single binary: will AI take a given job, or not. This paper argues that the accurate picture is a taxonomy rather than a binary. AI rarely replaces a job wholesale; more often it automates a bundle of tasks within a job, and the size of that bundle varies with task structure, sector, regulatory setting, and, distinctively in India's case, the economics of AI access itself. We propose a four-level framework (task, job, occupation, labour-market outcome) and classify Indian job impact into two broad regimes: jobs where AI has automated most of the skill bundle and the human is reduced largely to validation, and jobs where the human remains the primary driver and AI functions as a productivity tool. Each regime splits further into policy-relevant sub-types. Drawing on Indian evidence from business process outsourcing, manufacturing, MSMEs, sanitation, financial services, law, and software engineering, we show that every job in our sample has had to be reimagined, through task-level augmentation, task reallocation, or a genuinely new role built around operating the machine, rather than simply eliminated or defended unchanged. We close with a distributional analysis of who bears the cost of this reimagining and a three-pronged policy recommendation: continue encouraging AI diffusion, invest in India-specific labour-market research, and pair adoption with structured worker-transition pathways.

# Executive Summary

Public debate on AI and employment in India tends to ask a single question: Will AI take this job, or not? This white paper argues that the accurate picture is a **taxonomy rather than a binary**. AI rarely replaces a job wholesale; more often, it automates a bundle of tasks within a job. How large that bundle is, and whether it leads to augmentation, redesign or displacement, depends on task structure, sector, regulation, Responsible AI considerations and, distinctively in India, the economics of AI access itself.

## *Highlights*

1. **The task, not the job title, is the right unit of analysis.** The paper separates four levels (task, job, occupation and labour-market outcome) that public debate routinely conflates.
2. **India's AI-cost geography shapes what gets automated.** Cheap, "good-enough" AI diffuses quickly in BPO, bookkeeping, and MSME quality inspection, while frontier-dependent uses such as legal research and diagnostic imaging remain concentrated in urban, enterprise settings.
3. **Technically feasible automation can still fail commercially,** as quality-driven rollbacks of aggressive AI-first customer service show.
4. **The costs fall unevenly** across entry-level vs experienced, formal vs informal, urban vs rural, and MSME vs large-firm workers. The entry-level career pipeline is especially at risk.
5. **Every job affected by AI has to be reimagined,** whether through task-level augmentation, task reallocation, or new roles built around operating the machine, such as sanitation workers retrained as robot operators.

## *Two regimes, seven impact pathways*

| Regime | Impact pathway | Indian example |
|---|---|---|
| **1: AI performs most routine tasks; human validates** | **1a Cheap, "good-enough" AI** | Tier-1 BPO voice support; GST bookkeeping; Hyundai Chennai quality checks; TN MSME shop-floor tools |
| | **1b Frontier-dependent, costly** | Advanced legal research; CT/MRI radiology triage |
| | **1c Rollback** | Klarna-style reversal of over-automated customer service |
| | **1d Humans should not be there** | Robotic manhole and sewer cleaning (Bandicoot, Chennai) |
| **2: Human remains the primary driver; AI is a productivity tool** | **2a Human touch and judgment** | High-value loan and insurance sales; senior software engineering |
| | **2b Regulatory necessity** | Statutory audit sign-off; driver "effective control" under the Motor Vehicles Act |
| | **2c Capability gap** | Litigation advocacy; multi-issue, code-switched support calls |

## *New roles, not only subtraction*

AI is also creating occupations that barely existed before: Forward Deployed Engineers, AI trainers, ethical AI specialists, escalation specialists, AI conversation designers and remote robot operators.

## *Recommendations*

- **Encourage AI adoption and diffusion,** especially affordable, narrow, sector-specific tools for MSMEs.
- **Invest in India-specific AI and labour-market research** to map which skill bundles are safe to automate and which carry hidden quality, trust or safety costs.
- **Pair adoption with structured worker-transition pathways** and AI-awareness education for workers, firms and regulators, so liability and skilling frameworks are in place before adoption accelerates.

# 1. How AI Affects Work

## 1.1 A four-level framework

Discussions of AI and employment routinely conflate four distinct levels of analysis, which produces much of the confusion in the public debate. We define each level and use it consistently through the paper.

| Level | Working definition | Illustrative Indian example |
|---|---|---|
| Task | A single, describable unit of work with a definable input and output. | Verifying an OTP; reading a CT scan for a nodule; entering an invoice line into a ledger. |
| Job | A bundle of tasks routinely performed by one worker in one role. | A Tier-1 BPO agent's job bundles password resets, OTP verification, and order-status queries. |
| Occupation | A standardised classification of similar jobs across employers (e.g. NCO-2015 codes). | Customer service executive, quality inspector, chartered accountant, product designer. |
| Labour-market outcome | Aggregate change in employment, wages, or working conditions across an occupation or sector. | Net BPO hiring contraction in Tier-1 cities; growth in senior software-engineering demand. |

Most AI-and-jobs commentary jumps directly from task-level capability (“AI can now read a CT scan”) to labour-market-outcome claims (“radiologists are finished”), skipping the job and occupation levels where the actual redesign happens. This paper deliberately works through all four levels for each example.

A further consideration cuts across all four levels: the impact of AI on work cannot be separated from Responsible AI and governance. Whether a technically capable system should perform a task also depends on questions of safety, fairness, accountability, transparency and regulatory compliance. These considerations become particularly important as AI systems move from assisting humans towards making bounded decisions and taking actions across workflows. Understanding AI's impact on jobs therefore requires examining not only what AI can automate, but also what organisations can responsibly and lawfully delegate to it. This framing is consistent with the India AI Governance Guidelines released by MeitY under the IndiaAI Mission in November 2025, whose principles, including "people first", accountability, "understandable by design", and safety, resilience and sustainability, emphasise that decisions about delegating work to AI must account for human oversight and accountability, and not technical capability alone (MeitY, 2025).

One further distinction matters for the task-to-job argument: the scope of AI agency. AI may assist with an individual task, make a bounded decision, or execute a sequence of work across a workflow. As that scope expands, the residual human task bundle can change, but greater agency does not by itself imply greater job displacement; cost, capability, human value, safety, regulation and organisational design still determine the outcome.

## 1.2 Overview

The taxonomy developed here is an analytical framework informed by the authors’ experience, practitioner conversations and the cases and literature reviewed; the examples are illustrative rather than a representative sample of the Indian labour market. The paper therefore does not claim that the pathways identified here are exhaustive, universal, or distributed across occupations in the proportions suggested by the examples.

The impact pathways developed here explain why and under what conditions AI changes a job. The scope of AI agency is treated as a cross-cutting consideration rather than a separate employment taxonomy: outcomes depend on how agency interacts with cost, capability, human value, safety, regulation and organisational design.

Public debate on AI and employment tends to compress a genuinely complicated phenomenon into a single binary: will AI take this job or not. The more accurate picture, visible once actual deployments across Indian sectors are examined, is that AI more commonly automates selected tasks than entire occupations, although job- or occupation-level displacement can occur downstream when the residual human task bundle becomes too small, intermittent, or reorganised to sustain the original dedicated role (International Labour Organization, 2025). AI can perform or materially reshape a bundle of tasks within a job; whether that task-level change ultimately produces augmentation, role redesign, or displacement varies enormously by task, sector, organisational design, regulatory setting, and, in India's case, by the economics of AI access itself. Displacement is therefore treated here as a labour-market outcome of task and job redesign, not as a capability that follows automatically from the technology. A useful taxonomy divides jobs into two broad regimes: those where AI has automated a large share of the skill bundle and the human is reduced mostly to validation, and those where the human remains the primary driver and AI functions as a productivity tool. Each regime further splits into policy-relevant sub-types, and India's specific constraints on AI cost and access cut across all of them. The unifying conclusion, whichever regime a job falls into, is that jobs affected materially by AI may therefore need to be reimagined, not merely trimmed or defended unchanged, but rethought around a new division of labour between human and machine.

# 2. A Map of Job Types

Two dimensions determine how a task responds to AI: how structured the work is, and the kind of value the human provides. The table below illustrates the framework with representative Indian examples rather than attempting an exhaustive cross-classification of every task type.

| Structure | Human value provided | Indian example | Likely AI effect |
|---|---|---|---|
| Routine / rules-based | Information processing | OTP verification, GST invoice OCR | High: commodity models suffice |
| Routine / rules-based | Physical execution | Repetitive weld inspection | High, where robotics/vision is deployed |
| Semi-structured / exception-heavy | Judgment and accountability | Bookkeeping with exception review | Partial: human validates edge cases |
| Semi-structured / exception-heavy | Trust, empathy, persuasion | High-value insurance sales | Low-to-partial: conversion falls when AI leads |
| Non-routine / judgment-intensive | Relationship management and accountability | Litigation strategy, client counsel | Low: AI assists research, not judgment |
| Non-routine / judgment-intensive | Original creative expression | Advertising copywriting, graphic design, film/content editing, music production | Partial: AI drafts and generates variants; human retains authorship, taste, and final judgment |
| Non-routine / judgment-intensive | Safety-critical intervention | Manhole entry, hazardous-site work | High priority for automation on safety grounds, independent of cost |

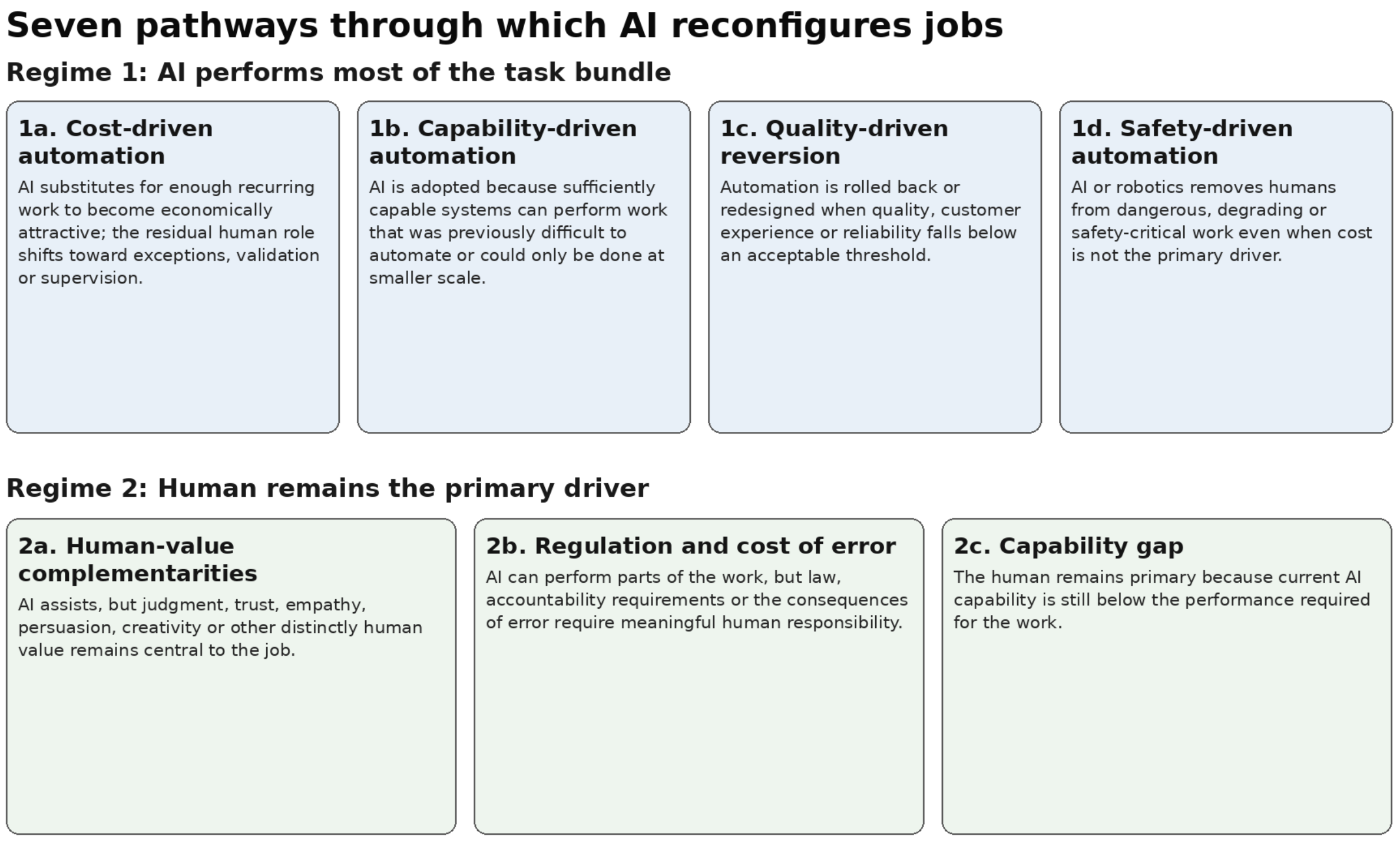

*Figure 1. Seven impact pathways through which AI can reconfigure jobs. Technical feasibility alone does not determine job outcomes; the effect of AI also depends on economic, capability, safety, human-value and regulatory conditions surrounding the work.*

A job may sit at any point along this continuum. Movement towards greater AI autonomy depends not only on technical capability, but also on cost, reliability, safety, regulation, human value and the consequences of error.

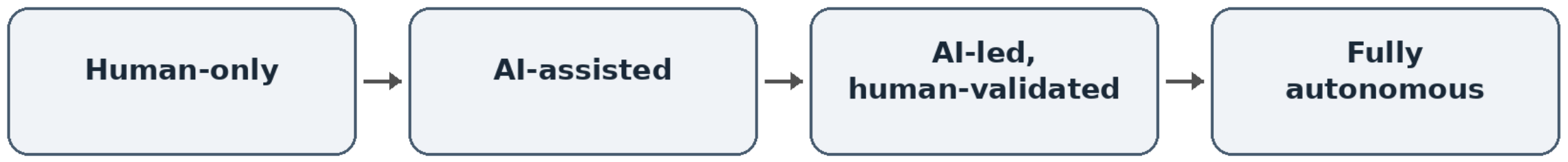

*Figure 2. Most jobs sit somewhere along this continuum rather than at either endpoint. Greater AI agency can shift the unit of automation from individual tasks and bounded decisions towards workflows and processes, but does not by itself determine where a job will sit on this continuum; the impact pathways in Section 3 help explain that outcome.*

# 3. Impact Pathways

The sub-types below are organised by the factor that most determines the outcome in each case, cost and capability (1a, 1b), organisational outcome (1c), ethical purpose (1d), human value (2a), regulation (2b), or technical capability (2c). This is a classification of impact pathways rather than a strict, mutually exclusive taxonomy: a single job can sit in more than one sub-type at once, for example, a job can be simultaneously frontier-dependent *and* legally required to retain human accountability *and* vulnerable to an automation rollback. Readers should treat the sub-type headings as lenses on a job rather than as a single correct label.

## Regime 1: AI performs most of the routine task bundle, human validates or handles exceptions

### *Sub-type 1a: Cheap, “good-enough” AI displaces at scale*

The clearest Indian example is Tier-1 voice support in the business process outsourcing (BPO) sector. Routine password resets, order-status queries, appointment confirmations, OTP verification calls, and payment reminders are increasingly handled by voice AI agents built on modest, non-frontier language models combined with speech recognition and text-to-speech, technology that does not require access to the most expensive models. The economic logic is well documented at the global level by Gartner, even where India-specific figures are thinner: labour represents up to 95% of contact-centre operating costs, and conversational AI deployments were forecast to cut global contact-centre agent labour costs by $80 billion in 2026 (Gartner, 2022), with newer, agentic AI systems expected to autonomously resolve 80% of common customer-service issues without human intervention by 2029, cutting operational costs by roughly 30% (Gartner, 2025). No single authoritative study publishes a matching India-specific cost breakdown, but Table 1 triangulates the figure across several independent industry pricing guides, which converge on a range broadly consistent with Gartner's global picture: a fully loaded Indian BPO seat costs roughly four to five times as much per minute as managed, production-grade voice AI, a gap wide enough that even mid-tier, non-frontier models are “cost-effective” today in the strict sense: cheaper to run than to pay a human, with no need for cutting-edge capability.

*Table 1. Fully loaded human BPO agent vs. voice AI cost, India, 2026 (triangulated across independent industry sources; no single authoritative study publishes this comparison directly).*

| Benchmark | ₹ / minute | Basis | Source(s) |
|---|---|---|---|
| Human BPO agent, fully loaded (direct INR estimate) | 8–20 | Salary, infrastructure, management, idle time; Tier-1/2 city | awaaz.ai 2026 |
| Human BPO agent, fully loaded (USD/hr, converted @ ≈₹85/$) | 8.5–23 | Offshore India/Philippines agent-hour rate, 4 independent guides | 1840 & Co.; Callforce Global; Crescendo.ai; Helpware 2026 |
| Voice AI, managed / production-grade | 4–9 | Blended platform + telephony + STT/TTS + LLM cost at volume | Caller Digital; Edesy Blog 2026 |
| Voice AI, self-serve API (cheapest tier) | 0.4–2 | Bare per-minute API rate; excludes integration, support, escalation | Scalify Labs 2026 |

Gartner cautions, however, that this arbitrage is not a fixed technical advantage but a function of current, partly subsidised AI pricing: it separately forecasts that by 2030, rising data-centre costs and the shift of AI vendors from subsidised growth toward profitability will push the cost per resolution for generative AI above $3, which would exceed the cost of many offshore human agents (Gartner, 2026), a genuine risk that today's cost gap narrows or reverses well before India's BPO sector has finished restructuring around it. At the 2026 World Economic Forum Annual Meeting, Harvard economist and former IMF First Deputy Managing Director Gita Gopinath warned that business process outsourcing “is going to be impacted by AI” (World Economic Forum, 2026). India's top IT firms added only 17 net employees in the first nine months of fiscal 2026, signalling a sharp contraction in entry-level demand (The Economic Times, 2026). Similarly, routine bookkeeping and data-entry tasks for Chartered Accountants, invoice OCR, bank reconciliation, GST return pre-filling, ledger classification, are increasingly handled by AI tools purpose-built for Indian compliance (GST, TDS, Tally integration), cutting manual bookkeeping time in reported vendor deployments, though independent, audited figures for the scale of this time saving are not yet available. In both cases the human's residual role is largely to validate: sign off on the CA's filings, or escalate the share of calls the AI cannot resolve.

As systems move from assisting with individual tasks towards bounded decisions and workflow-level execution, the relevant question shifts from how much of an individual task AI can perform to how much of the job's recurring workflow can be executed without continuous human intervention.

Tamil Nadu's manufacturing base shows the same logic on the shop floor. Hyundai's Sriperumbudur plant near Chennai has adopted AI-assisted visual scanning and digital validation across the vehicle-body assembly process, part of a broader Industry 4.0 and IIoT push at the plant (Business Standard, 2024). At the group level, Hyundai Motor Group and Kia's Digital Measuring Centre now validates around 1,000 measurement points on every vehicle body against roughly 600 quality-evaluation criteria, using optical 3D scanners and ultra-high-speed cameras rather than manual inspection (Automotive World, 2026), the standards from which are transferred into plant-level manufacturing processes across the group. This is not exotic AI: robotic inspection tools and rule-based checks for welding precision and sealant application are mature, inexpensive-to-run systems, well inside Sub-type 1a's cost-effective automation zone. The state government's own push reflects this pattern at MSME scale: Tamil Nadu allocated ₹398 crore to its AI, Information Technology and Digital Services Department in the 2026 state budget (dtnext.in, 2026; Deccan Chronicle, 2026), alongside the IIT-Madras–FaMe TN partnership, both aimed at building low-cost, user-friendly AI tools, anomaly detection, production-analytics dashboards, smart quality inspection, designed for small automotive, engineering, and fabrication units that cannot afford frontier-grade systems.

A similar cost gap is emerging in creative and design work, illustrating that Sub-type 1a's logic extends beyond manufacturing and BPO, though the evidence here is weaker and worth flagging as such. Industry pricing guides put a human graphic designer's rate in India at roughly ₹300–₹2,000 per hour depending on experience and city (Internshala, 2026), implying a typical single design asset (0.5–1.5 hours of work) costs on the order of ₹150–₹3,000. Neither Adobe nor Midjourney publishes a simple per-image price, so any INR/image figure is a derived estimate rather than an official rate. On Adobe Firefly, the Standard plan bills $9.99/month for 2,000 generative credits, and a standard-tier image generation typically draws 1 credit (Adobe, 2026a, 2026b; G2, 2026; Capterra, 2026), implying a cost of roughly ₹0.40–₹0.50 per standard image at ≈₹85/$; premium or high-resolution generations, which draw 10–20 credits each, cost roughly ₹4–₹8. On Midjourney, which bills for GPU compute time rather than per image, industry estimates derived from the Basic and Standard plans put a standard Fast-mode image at roughly ₹3–₹4 (eesel AI, 2026; TechJack Solutions, 2026; Nesyona, 2026), though Standard-tier and higher subscribers also get unlimited (slower) generation via “Relax Mode” at no extra cost, so a heavy user's effective per-image cost can run well below the Fast-mode figure. Taken together, a single AI-generated image costs on the order of a few tens of paise to a few rupees for standard output, rising to roughly ₹8–₹10 for premium or high-resolution generations, against roughly ₹150–₹3,000 for a comparable human-made asset.

*A caveat on this comparison specifically:* unlike the Table 1 BPO figures, which triangulate four independent India-focused cost guides against a direct INR estimate, the creative-cost figures above rest mostly on US-dollar-denominated SaaS pricing pages and third-party comparison blogs rather than an Indian survey or an official per-image rate card from Adobe or Midjourney; the INR conversion (≈₹85/$) will also drift with the exchange rate. The Indian designer-rate figure itself varies by source, Internshala's own site quotes ₹300–₹2,000/hour on one page and ₹2,000–₹12,000/hour for freelance

hires on another, reflecting real differences by experience and hiring channel rather than a single settled rate. These figures should be read as an illustrative order-of-magnitude comparison, not an audited benchmark.

### What narrow AI can already do for MSMEs, and what remains to be built

MSMEs are the sector where Sub-type 1a's logic matters most at scale: India has roughly 63 million MSMEs employing around 110 million people (IIFL, 2026) (a widely cited estimate; the government's Udyam registration portal recorded a smaller but fast-growing 7.47 crore *registered* enterprises employing 32.8 crore people as of the 2025–26 Economic Survey, reflecting the gap between total and formally registered MSMEs (Press Information Bureau, 2026a)), and almost none of them can afford frontier-model deployments; the entire opportunity depends on narrow, task-specific AI staying cheap. Several use cases are already commercially deployed by AI vendors serving Indian MSMEs: camera-based visual quality inspection using computer-vision models rather than generative AI; vibration- and temperature-sensor-based predictive maintenance flagging bearing wear or motor degradation before failure; AI-driven demand forecasting using regression or random-forest models trained on a firm's own sales history; and multilingual AI chatbots handling order status and FAQs for MSMEs selling on digital marketplaces. Vendors report substantial gains from these deployments, lower defect rates, reduced unplanned downtime, improved forecast accuracy, but these figures come from vendor case studies and pilots rather than independently audited, representative samples, and should be read as illustrative of what narrow AI can do rather than as sector-wide averages (see Section 7 on evidence quality). In the automotive-component cluster around Chennai specifically, AI-and-robotics-enabled Common Facility Centres are reported to be pooling machine-vision and production-optimisation tools that individual small units could not afford alone, directly lowering per-unit automation cost across the Precision Engineering Cluster.

A second layer of narrow-AI applications is technically feasible today but not yet widely deployed across Tamil Nadu's MSME base, representing the clearest near-term opportunity for policy-supported diffusion. AI-based credit assessment using alternative data (GST filings, UPI transaction history, e-commerce sales) could extend formal credit to MSMEs that lack the collateral or credit history conventional lenders require. Automated GST-linked expense-anomaly detection and cash-flow alerting could extend the bookkeeping tools already used by Chartered Accountants directly into MSME-owned dashboards. Route and logistics optimisation using heuristic or reinforcement-learning-inspired algorithms remains underused among smaller Tamil Nadu exporters and logistics-dependent units. AI-assisted generative design and yield optimisation for job-shop and fabrication units, suggesting material-efficient part layouts or optimal batch sequencing, is technically mature in larger factories but has not been packaged affordably for the textile, leather, and light-engineering MSME clusters that dominate Tamil Nadu's Tiruppur, Coimbatore, and Vellore belts. Building and subsidising exactly these narrow, sector-specific tools, rather than waiting for cheaper frontier-model access, is where an MSME-focused AI-diffusion strategy has the most immediate payoff, and it is precisely the gap the IIT-Madras–FaMe TN partnership and the state's ₹398-crore allocation are aimed at closing.

### Sub-type 1b: Only frontier-capability AI can do the task reliably, and it may be too costly

A second, less-discussed sub-type covers tasks where AI could perform most of the relevant task bundle, but only models at or near the current capability frontier with strong reasoning, multilingual nuance, or domain depth reach acceptable reliability, and running those models at scale is not yet cost-effective for many Indian use cases. Complex legal research and contract review is illustrative: AI-assisted legal-research tools have measurably reduced the paralegal and entry-associate research burden at large Indian law firms, but genuinely reliable case-law synthesis and litigation strategy still require frontier-level reasoning (see Tripathi et al., 2024 on measured accuracy and fairness limitations of LLMs on Indian legal tasks) that is expensive to deploy at scale across the thousands of smaller

Indian law practices. Advanced radiology triage, reading CT and MRI scans for early cancer or tuberculosis detection at specialist-level accuracy, is another case; the underlying AI models are highly capable, but Indian hospitals report that infrastructural constraints, lack of formal training, and cost of deployment, more than raw model performance, remain the binding constraint on adoption, especially outside metro tertiary-care centres. This is the sub-type India's AI policy conversation should worry about most, because it is where the "frontier tax", the price premium of state-of-the-art capability over commodity models, most directly determines whether citizens outside big cities get access to AI-augmented diagnostics, legal aid, or financial advice at all.

### *Sub-type 1c: The rollback, when "replace aggressively" fails*

A cautionary sub-type shows that even where a large share of a job's skills look automatable, aggressive full replacement can backfire. In February 2024, Klarna, the Swedish fintech, announced jointly with OpenAI that a newly deployed AI assistant had handled 2.3 million customer-service conversations in its first month, a majority of the company's chat volume, doing what Klarna itself described as the equivalent work of 700 full-time agents; these figures are Klarna's own company-reported numbers, not an independently audited count (Klarna, 2024). By May 2025, CEO Sebastian Siemiatkowski told Bloomberg that the cost-driven push had gone too far and produced "lower quality," and Klarna began recruiting human agents again, under a flexible, remote "Uber-style" model, for disputes, complex cases, and any customer who wants the option of a human, while keeping AI on routine volume (Bloomberg, 2025; The Independent, 2025). The lesson generalises directly to Indian BPOs and financial services now running similarly aggressive playbooks: the reliably automatable share of a job can be handed to AI, but the remaining share, de-escalation, empathy, multi-issue conversations, trust-building on money decisions, breaks the model if pushed too far, too fast. Indian banks and telecom operators piloting full AI-first support should treat Klarna as a live case study in the limits of Regime 1 logic (see also Mohanty (2026) for a retail-banking illustration of how failures in agentic systems can cascade from individual harm to business disruption and wider systemic risk).

### *Sub-type 1d: Jobs where humans should not be there at all*

A distinct and morally urgent sub-type covers occupations where the goal of automation is not cost efficiency but the removal of humans from lethal conditions, sanitation work is the starkest case. Manual scavenging, the practice of entering manholes and septic tanks by hand to clear blockages, has been illegal in India since 2013, yet government data placed before the Lok Sabha in August 2026 recorded 498 sanitation-worker deaths during sewer and septic-tank cleaning between January 2019 and June 2026, with the government maintaining that none of these deaths is classified as a manual-scavenging death, a classification gap that advocacy groups argue understates the true toll (Business Standard, 2026b). Robotic systems such as Genrobotics' Bandicoot are explicitly engineered to make human entry into manholes unnecessary, using robotic arms, gas sensors, AI-based anomaly detection and camera vision to replicate a scavenger's actions from outside the manhole (The Better India, 2024–2025; YourStory, 2024–2025). The clearest documented Indian deployment is Chennai's: the Chennai Metropolitan Water Supply and Sewerage Board, a government body, has operated Bandicoot Mobility+ robots since April 2024 and reported that in their first year the robots cleaned over 5,000 manholes and stormwater drains across the city (Machine Maker, 2025; Construction World, 2025; DQ India, 2025). This sub-type inverts the usual cost debate: the AI system is typically more expensive up front than paying human labour, but the case for automation rests on eliminating preventable deaths and constitutional-dignity violations rather than on cost-effectiveness, and the residual human role shifts toward remote robot operation and equipment maintenance. The Ministry of Social Justice and Empowerment's NAMASTE programme has profiled and validated tens of thousands of sewer and septic-tank workers nationally and is oriented around eliminating hazardous manual entry (Press Information Bureau, 2026b; ETV Bharat, 2025), though a public, independently verified count of how many workers have specifically been retrained as robot operators under this or similar programmes is not available, and we do not assert a figure here. Municipal deployment nonetheless remains patchy nationally, illustrating that even in the one category where nobody disputes AI should do the work, adoption still lags for financing and operator-training reasons rather than technical ones.

## Regime 2: Human remains primary driver, AI is a productivity tool

### *Sub-type 2a: Human judgment and "the human touch"*

Here AI meaningfully accelerates the job without displacing its core, because the value proposition depends on human presence itself. Industry practitioners in Indian financial services report that outbound sales for high-value products, large-ticket loans, high-premium insurance, see conversion rates fall when AI voice agents replace humans, because customers, particularly older ones, still want a human voice for major money decisions; precise, independently audited conversion-decline figures are not publicly available and the direction, rather than the exact magnitude, is the more defensible claim. Collections and recovery calls show a similar pattern: negotiation and empathy remain firmly human work. Senior software engineering shows a parallel dynamic, AI coding assistants and coding agents (such as GitHub Copilot and Claude Code) now write a substantial share of routine, boilerplate-heavy code and have compressed entry-level, boilerplate roles, while system architecture, debugging of complex failures, and technical leadership remain squarely human because they require judgment AI does not yet reliably supply. Reported industry figures for the exact share of AI-authored code and for demand growth in senior engineering roles vary considerably by source and methodology; we flag them here as directionally consistent with practitioner reporting rather than as precisely measured statistics (see Section 7).

### *Sub-type 2b: Regulatory necessity and the cost of AI error*

A second sub-type keeps humans "in the loop" not because AI cannot technically do the work, but because the law requires human accountability or because AI failure is too costly to permit full automation. "Not good enough" in this context need not mean technical underperformance alone. An AI system may be unsuitable for autonomous deployment because it fails to meet human-set ethical or safety requirements, does not comply with applicable regulation, lacks sufficient resilience for the autonomy being granted to it, or cannot provide the auditability and accountability required for consequential decisions. Greater AI agency does not necessarily imply greater scope for human removal: in regulated or consequential settings, movement from bounded decisions towards workflow-level execution may instead increase the importance of explicit human approval points, escalation rules and accountability. India's autonomous-vehicle law is the clean example: the Motor Vehicles Act, 1988 requires that a motor vehicle remain under the "effective control" of its driver at all times. The Act predates SAE automation-level terminology and does not name specific automation levels, but legal commentators note that its driver-centric, effective-control requirement effectively forecloses fully driverless (SAE Level 4–5) operation on Indian roads under the current statute, since India has no separate statutory framework assigning liability to a software system in the absence of a human driver. Statutory audit sign-off in accounting is a comparable case, AI can draft the numbers, but only a human Chartered Accountant can certify financial statements under ICAI rules, because the professional and legal liability for a wrong certification is severe.

### *Sub-type 2c: AI hasn't yet matched human-level performance*

The third sub-type is simply capability-limited: AI's ceiling in the task, today, sits below reliable human performance for the harder parts of the job. In law, litigation advocacy, client counsel, and courtroom advocacy remain beyond AI, since these responsibilities involve litigating before judges and giving clients real-time assistance in ways that require situational judgment models do not yet possess. In software, complex multi-system debugging and novel architecture design under ambiguous requirements remain difficult for AI to execute reliably without heavy human direction. In customer service, industry practitioners report that multi-issue conversations combining billing, complaint, and product questions in a single call, and regional-language code-switching (Hindi-English or Tamil-English mixing with emotional nuance), remain areas where Indian-language AI underperforms humans in real-world conditions, a pattern consistent with practitioner experience though not yet the subject of a rigorous, published benchmark specific to Indian code-switched speech.

Creative industries add a further illustration of this Regime 2 logic, visible at global scale even though India-specific data is not yet available. Adobe's own disclosures show its Firefly generative tool had produced more than 22 billion assets worldwide by April 2025 (Adobe Blog, 2025), and the company reported AI-influenced annual recurring revenue surpassing $5 billion by the third quarter of fiscal year 2025, reflecting how deeply generative AI has been embedded into professional creative workflows such as scriptwriting for short commercials, photography, and graphic design in a little over two years (Adobe Inc., 2025; SQ Magazine, 2026; Skilldemia, 2026). A Forrester Total Economic Impact study commissioned by Adobe found that generative AI tools boosted creative-ideation productivity by 30% to 70% among the enterprise creative teams it studied, while final creative judgment, shaping narrative nuance, cultural context, and brand voice, remained with human creatives throughout (Forrester Consulting, 2025). That finding should be read with its methodology in mind: it is a vendor-commissioned study based on eleven interviewees across six organisations modelled into a single hypothetical composite organisation, rather than an independent or India-specific survey, and its productivity range describes risk-adjusted projections rather than a measured industry average. A similar pattern is reported in UX design, where practitioner guidance frames AI as best used as a co-pilot for secondary research, persona exploration, and routine interface-layout work, with a human-in-the-loop retaining final say on accuracy, intent, experience quality, and, in particular, empathy and critical thinking (Maze, 2025). None of this evidence is India-specific, but it corroborates, at global scale, the same pattern of AI accelerating without displacing core judgment that Sub-type 2a documents in Indian sales and engineering roles above.

# 4. Why India's AI-Cost Geography Changes the Calculus

The cost-efficiency dimension is not incidental to this taxonomy, in India it often determines which sub-type a job falls into. India currently lacks a mature, general-purpose sovereign frontier model comparable to global closed models; building one from scratch requires cluster-months on many thousands of top-end GPUs and outlays that only a handful of global labs have funded. Krutrim's attempt to build frontier capability, custom silicon, and a consumer app simultaneously on its own balance sheet illustrates the difficulty: by May 2026 the company had paused its in-house chip-design effort and shifted from AI model development to an AI-cloud business, as its model-building ambitions met commercial reality (TechCrunch, 2026; Business Standard, 2026a). Sarvam AI's approach, training a 105-billion-parameter open-weight model on subsidised IndiaAI Mission compute (a ₹246.72-crore compute allocation, roughly 40% of it government-subsidised, and access to 4,096 NVIDIA H100 GPUs over roughly six months) (NVIDIA, 2026; Forbes, 2026; niftytrader.in, 2026), , shows a viable middle path, pricing inference at $0.80 per million blended tokens against roughly $4.50 for OpenAI's GPT-5.4 Mini, a smaller closed model, an 82% discount (Free Press Journal, 2026). Twenty sovereign model proposals (12 large language models and 8 small language models) have been identified for support under the IndiaAI Mission as of July 2026 (Business Standard, 2026), but most remain narrower or cheaper than the true frontier, which is exactly why India's AI adoption skews toward the “cheap, good-enough” automation of Sub-type 1a far more than the frontier-dependent automation of Sub-type 1b.

India's AI-cost geography can also be understood through the broader lens of Frugal AI. There is no single settled definition of Frugal AI; for the purposes of this paper, we use frugality primarily in the engineering sense of sufficiency and resource optimisation: achieving the functionality, reliability and scale required for a task while minimising unnecessary demands on compute, data, energy, infrastructure and cost. Frugality therefore need not mean simply choosing the cheapest or smallest model. A frontier-capability system could also be frugally engineered where that level of capability is necessary and resources are used efficiently. In India, where deployment must accommodate price sensitivity, uneven compute and connectivity, energy constraints and very large-scale use, such frugality can become an engineering requirement rather than merely a cost-saving preference. This conception has roots in a longer Indian engineering tradition. Research at IIT Madras has treated frugality not as improvised "jugaad" or low cost alone, but as a rigorous engineering problem: conserving resources and cost while maintaining robust functionality.(Rao, 2019; India Today, 2026) Greater efficiency may, however, also increase overall AI use, creating a possible rebound effect: lower resource use per task need not translate into lower aggregate resource demand as deployment expands.(International Energy Agency, 2025; International Energy Agency, 2026).

This has a direct bearing on the "AI for every citizen" ambition. National Sample Survey data present a mixed picture on rural access: as of the 2022–23 round, only around 4% of rural households owned a computer or laptop (versus roughly 22% in urban areas) (Data For India, n.d.), which does constrain AI-based learning and service access that depends on a full computing device. Smartphone penetration, however, has grown much faster than computer ownership and is considerably higher than often assumed: the National Sample Survey's 2025 telecom module found that roughly 82% of rural households owned a smartphone (against about 91% in urban areas), with 85.5% of all Indian households owning at least one smartphone (Ministry of Statistics and Programme Implementation, 2025). The binding constraint on rural AI access, in other words, is less raw device ownership than the quality of the device (feature phones versus smartphones), data affordability, and the small-screen, low-bandwidth suitability of the AI service itself, a materially different policy problem than the low-double-digit ownership figures sometimes cited in this debate. Commercial AI service providers in India, competing on thin margins against subsidised sovereign alternatives and price-sensitive customers, have limited financial headroom to subsidise frontier-model access for underserved segments, reinforcing a two-tier outcome: cheap automation diffuses fast in Sub-type 1a use cases (BPO, basic bookkeeping), while frontier-dependent uses (advanced legal research, sophisticated diagnostic imaging) remain concentrated in well-resourced urban and enterprise settings.

# 5. Taxonomy at a Glance

| Regime | Sub-type | Example (India) | Driving factor | AI sophistication needed |
|---|---|---|---|---|
| 1: AI performs most routine tasks | 1a: cheap, good-enough | Tier-1 BPO calls, GST bookkeeping, Hyundai Chennai quality checks, TN MSME shop-floor tools | Cost arbitrage | Commodity / narrow models |
| 1: AI performs most routine tasks | 1b: frontier-dependent, costly | Advanced legal research, radiology triage | Reliability requires frontier reasoning | Frontier models, expensive at scale |
| 1: AI performs most routine tasks | 1c: rollback | Klarna-style customer service reversal | Quality collapse on complex cases | Any: over-automation risk |
| 1: AI performs most routine tasks | 1d: humans should not be there | Manhole/sewer cleaning robots (Bandicoot) | Eliminate death and dignity risk, not cost | Narrow robotics + AI sensing |
| 2: Human remains driver | 2a: human touch/judgment | High-value loan sales, senior engineering | Trust, empathy, architectural judgment | AI as assistant or bounded agent; human remains primary |
| 2: Human remains driver | 2b: regulatory necessity | Autonomous vehicles, audit sign-off | Legal liability regime | Irrelevant: law mandates human |
| 2: Human remains driver | 2c: capability gap | Litigation advocacy, multi-issue support calls | AI ceiling below human floor | Current AI insufficient |

The scope of AI agency provides an additional lens on these job types. As AI moves from assisting with individual tasks towards bounded decisions and workflow execution, it can change the size and composition of the residual human task bundle. The employment outcome, however, continues to depend on the impact pathways above rather than on AI agency alone.

# 6. Who Gains and Who Loses

The taxonomy in Sections 3–5 describes *which* tasks and jobs change; it says less about *who* bears the cost of that change. A distributional lens is necessary because the same technology can raise productivity for one group while eroding the bargaining power of another, and because aggregate job-count statistics can mask a “missing middle” in which junior roles disappear even as demand for senior workers holds or grows, the senior-software-engineering pattern described in Sub-type 2a is one instance of this.

Several fault lines recur across the sub-types documented in this paper. **Entry-level versus experienced workers:** Sub-types 1a and 2a both compress or eliminate the routine, boilerplate-heavy tasks that historically gave junior workers their first foothold, while demand for experienced judgment holds or rises, a dynamic explored further in Section 6.2. **Formal versus informal workers:** the taxonomy in this paper is built almost entirely from formal-sector evidence (registered BPOs, listed manufacturers, chartered accountants, law firms); Section 6.3 addresses the much larger informal workforce separately, because AI's effect there runs through prices and platforms rather than direct redundancy. **Urban versus rural workers:** the cost-geography analysis in Section 4 shows that frontier-dependent automation (Sub-type 1b) concentrates in metro, enterprise settings, while narrow, subsidised automation (Sub-type 1a) is the more realistic near-term reach into MSME and rural contexts. **English-speaking versus regional-language workers:** Sub-type 2c suggests Indian-language AI still underperforms English-language AI on nuanced, code-switched conversation, which may temporarily protect regional-language customer-facing roles even as English-medium equivalents automate faster. **Large firms versus MSMEs:** large firms can absorb the fixed cost of frontier-model integration and in-house customisation; MSMEs depend on the narrow, subsidised tools described in Section 3, and the gap between the two is itself a policy variable, not a fixed feature of the technology. **Employees versus contractors and gig workers:** several of the sub-types documented here (Sub-type 1d's robot operators, Sub-type 1a's escalation specialists) are new roles created by automation, but whether they are offered as stable employment or piecework affects how much of the productivity gain reaches the worker, and this paper's evidence base does not extend to systematically tracking that split. **Credentialed versus experience-based workers:** AI-assisted tools can substitute for some forms of formal credentialing (e.g. AI-assisted bookkeeping reduces reliance on trained junior accountants for routine entries) while simultaneously raising the premium on credentials that carry statutory authority (Sub-type 2b's audit sign-off).

## 6.1 Job quality, not just job quantity

A change in job *quality* is distinct from a change in job creation or job loss, and the taxonomy in this paper is largely silent on it because the underlying Indian evidence is thin. As AI spreads further into Indian workplaces, several dimensions are worth tracking:

- **Wages:** do wages for AI-augmented roles rise, fall, or stay flat compared to the pre-AI version of the job?
- **Monitoring and surveillance:** does AI introduce new forms of performance tracking, such as AI scoring of a BPO agent's remaining human-handled calls, or dashboard-based productivity tracking for MSME shop-floor workers?

- **Work intensity and shifts:** do work intensity or shift patterns change as AI absorbs the lower-intensity part of a job, leaving humans with a residual stream of harder, more draining exceptions?
- **Stress and burnout:** what are the effects on stress and burnout, especially in escalation-heavy roles where humans increasingly handle only the calls the AI could not resolve?
- **Career progression:** how does career progression change once entry-level tasks are automated (addressed directly in Section 6.2)?
- **Quality of human interaction:** what is the quality of human interaction that customers and citizens actually receive?
- **Responsibility for errors:** who bears responsibility when an AI-assisted decision turns out to be wrong? Sub-type 2b's regulatory logic answers this for audit and driving, but the question remains open for most Sub-type 1a and 2a roles; structured AI incident reporting is one mechanism for surfacing such failures (Raju and Ravindran, 2025).

None of these dimensions is systematically measured in the Indian sources this paper draws on, and we flag that gap explicitly rather than assert a direction.

## 6.2 The entry-level career pipeline

Junior roles have historically functioned as an apprenticeship: repetitive, process-oriented tasks that are individually low-value but that collectively teach the judgment a worker later needs in a senior role. If Sub-types 1a and 2a automate precisely those apprenticeship-level tasks, password resets, basic contract review, boilerplate code, the open question is where the judgment required for senior roles will come from once the traditional training ground has shrunk. This is particularly relevant to India's IT, BPO, finance, and professional-services sectors, which have historically absorbed very large annual cohorts of entry-level graduates. One plausible response, illustrated for the BPO sector, is to redesign the entry point around supervising and correcting AI rather than performing the routine task directly, for example, an illustrative pathway such as Tier-1 voice agent → escalation specialist → quality evaluator → AI conversation designer → operations supervisor, in which each step still builds toward senior judgment, but the raw material the junior worker practises on is AI output rather than raw customer calls. Whether firms will actually invest in redesigning career ladders this way, rather than simply hiring fewer juniors and hoping senior judgment appears from elsewhere, is an open empirical question this paper does not resolve.

## 6.3 Informal employment

The taxonomy developed in Sections 3–5 is built almost entirely from formal-sector evidence, yet informal work, street commerce, small retail, domestic work, construction, agriculture, transport and logistics, self-employed professionals, and platform or gig work, employs a substantially larger share of India's workforce than the formal jobs this paper documents in detail. AI's effect on this much larger informal segment is likely to occur indirectly rather than through the kind of task-level redundancy described in Sections 3–4: through prices (AI-driven efficiency in formal supply chains changing what informal retailers can charge or source), through demand (automation-driven job losses in formal BPO and IT reducing local consumer spending that supports informal service providers), through credit access (AI-based alternative-data credit scoring, discussed in Section 3, could extend formal credit to informal-sector workers who currently cannot access it, or could exclude those without a digital transaction footprint), through platform mediation (gig and platform workers increasingly interact with AI-driven dispatch, pricing, and rating systems that shape their earnings without the platform itself “automating their job” in the Section 3 sense), and through business survival (a small retailer's livelihood depends on whether nearby formal businesses remain viable, not on whether the retailer's own tasks are automated). CeRAI’s study of India’s gig economy finds that algorithmic management already shapes work allocation and performance management for location-based platform workers, many of whom believe that a human, rather than an algorithm, is making these decisions (Kumar and Narayanan, 2025). This paper does not attempt to quantify these indirect channels; we flag informal employment as a priority for dedicated future research rather than force it into the direct-automation framework built for formal jobs.

# 7. Evidence and Limitations

This paper draws on a mix of government data, company disclosures, industry vendor reports, and journalistic accounts, which vary considerably in reliability. Throughout the paper we have tried to distinguish observed, measured deployments from pilots, vendor claims, and forecasts, and we summarise that distinction here using four labels.

| Label | What it means | Examples in this paper |
|---|---|---|
| Measured | Government data, audited disclosure, or independently reported figures. | Sarvam's ₹246.72-crore compute allocation. |
| Pilot / vendor claim | Reported by the deploying company or vendor without independent audit, often from a single site or small sample. | MSME visual-inspection, predictive-maintenance, and demand-forecasting figures in Section 3; Hyundai/Kia measurement-point figures reported at group level. |
| Industry estimate | Reported by industry practitioners or trade press, directionally consistent across sources but not from a single authoritative dataset. | BPO cost-per-minute ranges; conversion-rate declines for AI-led sales; India senior-engineering demand growth; Cognizant-Pearson 37% vs 33% entry-level task finding (survey of 750 HR leaders). |
| Forecast | A stated projection of a future state, not a current measurement. | Gartner's 80%-by-2029 customer-service automation forecast. |

Two limitations follow from this evidence base. First, India has comparatively little rigorous, published labour-market research tracking which specific skill bundles within specific occupations are actually being automated versus augmented, as distinct from employer or vendor self-report; this paper's sub-type classification should be read as a structured hypothesis for future empirical testing, not a settled finding. Second, several sub-type examples (particularly within 1b and 2c) rely on qualitative, practitioner-reported patterns rather than quantified benchmarks specific to the Indian context, and we have flagged those instances individually in Section 3.

# 8. Policy and Management Implications

The evidence supports encouraging AI diffusion in India but resists any single blanket prescription, because the “right” level of automation differs by sub-type and is often not yet known with confidence. Indian firms frequently do not know, in advance, whether a given job sits in Sub-type 1a (safe to automate heavily) or 1c/2a (where over-automation risks a costly Klarna-style reversal), this is as much an empirical gap as a technical one, since India has comparatively little rigorous labour-market research tracking which specific skill bundles within specific occupations are actually being automated versus augmented. The Cognizant-Pearson 2026 finding, based on a survey of 750 HR leaders, that 37% of entry-level tasks in India are already AI-performed, compared with 33% globally (Cognizant and Pearson, 2026), suggests India-specific dynamics, labour cost structures, language diversity, regulatory settings, are shaping outcomes differently than in the US or Europe, reinforcing the case for India-specific empirical study rather than importing conclusions from Western AI-labour research.

What the manufacturing, MSME, and sanitation examples add to this picture is the clearest illustration of the article's central claim: every job has to be reimagined, whether the reimagining ends in wholesale replacement, task-level augmentation, a smaller or redistributed human task bundle, or a completely new job category built around operating the machine. The Hyundai quality inspector's job was not eliminated by AI-assisted scanning, it was reimagined around interpreting flagged defects rather than performing the full visual check manually. The Tamil Nadu MSME fabrication worker's job is being reimagined around shop-floor dashboards and anomaly-detection alerts that a small unit could never have built in-house. Most strikingly, the sanitation worker's job has been reimagined from entering a manhole by hand to remotely operating a robotic arm from outside it, the underlying human function of keeping sewers clear survives, but the skill bundle required to perform it has been replaced almost entirely, from manual dexterity and risk tolerance to remote-control literacy and equipment maintenance. None of these outcomes fit a simple “AI took the job” or “AI didn't touch the job” narrative; each required deliberate redesign of what the job actually consists of. A dedicated job can cease to be viable even when some human tasks remain, if AI absorbs enough recurring work that the residual tasks are better organised as intermittent specialist review, supervision, escalation, or accountability work. This mechanism is distinct from simple technical substitution.

This reinforces the case for a three-pronged approach:

- Encourage AI adoption and diffusion, given its demonstrated productivity gains in the deployments documented above.
- Invest in dedicated AI, labour-market research to map, sector by sector, which skill bundles are genuinely safe to automate and which carry hidden quality, trust, or safety costs.
- Pair both with AI-awareness and adoption education:
  - For workers navigating role transitions (for example, retraining sanitation workers as robot operators, and MSME staff as dashboard-literate technicians).
  - For firms deciding where automation is cost-effective versus premature.
  - For regulators calibrating liability frameworks (as with autonomous vehicles) before, not after, adoption accelerates.

Framing the challenge as one where every job has to be reimagined, rather than one where jobs simply survive or disappear, should anchor India's AI-and-labour policy conversation going forward.

## Concrete worker-transition pathways

Any transition strategy needs to answer five practical questions for each affected worker group: which workers are displaced from which specific tasks; what adjacent roles they can realistically enter given their existing skills; what training duration is required to bridge the gap; who pays for that training, the employer, the state, or the worker; whether formal certification is necessary for the new role or whether on-the-job supervision suffices; whether wages in the adjacent role are preserved, and if not, how the gap is cushioned; and how workers are supported, financially and otherwise, during the transition period itself. The sanitation-to-robot-operator pathway under NAMASTE and the illustrative Tier-1-voice-agent-to-operations-supervisor pathway described in Section 6.2 are two concrete examples where such a pathway already exists in some form; most of the sub-types documented in Section 3 do not yet have an equivalent, and building one is a natural next step for both firm-level HR policy and state skilling programmes.

# 9. New Jobs AI Is Creating in India

Sections 3–9 focus on how AI displaces, augments, or reallocates tasks within existing jobs. That framing risks leaving the impression that AI's net effect on Indian employment is subtractive. It is not only that. The same transition is also creating occupations that did not exist, or existed only in small numbers, before AI adoption accelerated, and a credible account of AI's impact on Indian jobs has to include this side of the ledger.

The same reimagining can also create new roles around AI deployment and infrastructure, including Forward Deployed Engineers (FDEs), who bridge AI capabilities with operational workflows, and AI data-centre energy and infrastructure management roles.

The clearest official evidence comes from NITI Aayog's *Roadmap for Job Creation in the AI Economy*, developed with NASSCOM and BCG and released in October 2025. Alongside its warning that India's tech and customer-experience workforce faces contraction in routine roles, the roadmap projects that the same transition could generate up to 4 million new AI-related jobs over five years, and it explicitly names emerging Indian role categories, including Ethical AI Specialists, AI Trainers, Sentiment Analysts, and AI DevOps Engineers, that barely existed as job titles a few years earlier (NITI Aayog, 2025). The roadmap recommends a National AI Talent Mission to help realise this, and is explicit that the 4-million figure is an opportunity contingent on coordinated investment in skilling and reskilling, not an automatic outcome of AI adoption (NITI Aayog, 2025).

Several of the sub-types documented earlier in this paper illustrate the same pattern concretely rather than abstractly. Sanitation robotics (Sub-type 1d) has created a remote robot-operator role that did not exist before robotic deployment, and India's NAMASTE programme is oriented around building exactly this kind of transition for sewer and septic-tank workers, even though the precise number of workers who have moved into such roles nationally is not publicly documented (Press Information Bureau, 2026b; Genrobotics, 2026). MSME shop-floor staff are increasingly trained as dashboard-literate technicians who interpret anomaly-detection alerts rather than perform manual inspection alone (Section 3), a role with no direct pre-AI equivalent. The illustrative BPO career pathway described in Section 6.2, escalation specialist, quality evaluator, AI conversation designer, operations supervisor, names roles built specifically around supervising, correcting, and designing AI behaviour, several of which have only emerged as recognisable job titles in the past few years. And India's IndiaAI Mission compute allocations (Section 4) have created a small but growing cluster of AI infrastructure, data, and model-training roles at organisations such as Sarvam AI, alongside the broader AI-talent shift NITI Aayog documents at the national level (NITI Aayog, 2025; NVIDIA, 2026; Forbes, 2026; niftytrader.in, 2026).

None of this guarantees a positive net outcome for any individual displaced worker, and the paper does not claim it does: NITI Aayog's own roadmap frames new-job creation as conditional on “urgency, vision, and coordination” (NITI Aayog, 2025) rather than as something that happens automatically alongside displacement, and this paper's evidence base (Section 7) does not extend to systematically counting how many of the newly created roles are absorbing the specific workers displaced from Sub-type 1a and 2a jobs elsewhere in this taxonomy. But an account of AI's impact on Indian jobs that only counts subtraction is incomplete, and the new-role evidence above, official and sector-level, is the appropriate one.

# 10. Conclusion

This paper set out to move beyond the simple question of “Will AI take this job?” to a more useful one: which tasks are being performed or reshaped by AI in this particular case, when does that translate into automation or displacement, and how should the job be redesigned because of it? Across the sub-types documented here, five recurring reasons emerge: cutting costs (Sub-type 1a), reaching capabilities that are only possible with frontier-grade AI (Sub-type 1b), meeting regulatory requirements (Sub-type 2b), improving safety by removing humans from dangerous work (Sub-type 1d), and fixing service quality after an earlier automation attempt failed (Sub-type 1c).

Each reason calls for a different response. Cost-driven automation needs support for workers in transition, and vigilance against over-automation. Capability-driven automation needs investment to make frontier-grade AI more affordable, not just more available. Regulation-driven automation needs clear liability rules set up in advance, rather than left to catch up after deployment. Safety-driven automation needs funding that does not hinge on cost savings. And quality-driven rollbacks need humility about how far automation can be pushed before it breaks.

For India, this means employers, workers, and policymakers should stop treating every instance of “AI affecting jobs” as the same story. As AI moves from assistance on individual tasks towards bounded decisions and workflow-level execution, the unit of change may itself shift from the task towards the workflow; whether this ultimately displaces, augments or redesigns a job will still depend on the mechanisms identified in this paper. This also raises a longer-term question about the reproduction of expertise: if AI absorbs much of the entry-level work through which people historically acquire domain knowledge and judgment, how will organisations develop the next generation of human reviewers, supervisors and experts? Multiple AI stories are unfolding across the Indian labour market at once, and the right response depends on knowing which one is in front of you.

This paper is intended as the first step in a broader programme of work on AI and employment in India. The taxonomy developed here provides a set of hypotheses and impact pathways to be tested and refined through subsequent research, including systematic evidence reviews, sector- and work-profile analysis, and primary field research with employers and workers. Such work can examine how widely these pathways occur, how they vary across sectors and occupations, including areas such as education and healthcare, and how the scope of AI agency interacts with the job-change mechanisms identified here across sectors and occupations. This work will also connect with CeRAI's ongoing research on Primers and Grounding Responsible AI principles project and the perception study with CMC, to examine how deployment, governance and public perceptions interact with changes in work. Future work in this series will also draw on CeRAI's ongoing research on AI governance and public perceptions.